\documentclass[prl,aps,amssymb,amsfonts,amsmath,showpacs,superscriptaddress,twocolumn]{revtex4}
\usepackage{graphicx, epstopdf} % Einbindung von Grafikdateien
\usepackage{amsmath} %%% math symbols needed
\usepackage{amsfonts}%%% math fonts needed

\newcommand{\mplaffil}{Max Planck Institute for the Science of Light and Friedrich-Alexander-Universit\"at Erlangen-N\"urnberg (FAU), D-91058 Erlangen, Germany}
\newcommand{\lensaffil}{National Institute of Optics (INO-CNR) and European Laboratory for Nonlinear Spectroscopy (LENS)
via Nello Carrara 1, 50019 Sesto Fiorentino (FI), Italy}

\begin{document} \title{Coherent interaction of a metallic structure with a single \\
quantum emitter: from super absorption to cloaking }
\author{Xue-Wen Chen}
\affiliation{\mplaffil}
\author{Vahid Sandoghdar}
\affiliation{\mplaffil}
\author{Mario Agio}
\affiliation{\lensaffil}

\begin{abstract} We provide a general theoretical platform based on quantized radiation in absorptive and
inhomogeneous media for investigating the coherent interaction of light with metallic structures in the
immediate vicinity of quantum emitters. In the case of a very small metallic cluster, we demonstrate
extreme regimes where a single emitter can either counteract or enhance particle absorption by three
orders of magnitude. For larger structures, we show that an emitter can eliminate both scattering and
absorption and cloak a plasmonic antenna. We provide physical interpretations of our results and
discuss their applications in active metamaterials and quantum plasmonics. \end{abstract}

\maketitle
\textit{Introduction} - An isolated quantum mechanical two-level atom has extinction cross section $\sigma_{0}=\frac{3\lambda^2}{2\pi\epsilon_{b}}$ on resonance and can block an incoming light beam, where $\lambda$ is the vacuum transition wavelength and $\epsilon_{b}$ is the relative dielectric constant of the background medium~\cite{Zumofen08}. For a metallic structure made of a large number of atoms, the cross sections for scattering ($\sigma_{sca}$) and absorbing ($\sigma_{abs}$) light can be obtained from classical calculations and depend on the real and imaginary parts of its dielectric function as well as its size and shape~\cite{Bohren83}. While objects much smaller than the wavelength do not affect propagation of a light beam, we know from everyday experience that large metallic structures can fully extinguish it. The past several decades have witnessed a great interest in the effect of metals on the spectral properties of atoms and molecules~\cite{Moskovits85}. Theoretical treatments of these phenomena usually consider the absorption and emission of a classical dipole close to a metallic particle, which acts as an optical antenna to modify the spontaneous emission, open new nonradiative channels or enhance the Raman cross section~\cite{Agio13}.

Recently, several reports have considered the \textit{coherent} interaction of an incident light beam with a semiconductor quantum dot coupled to a very small metallic nanoparticle~\cite{Zhang06,Artuso08,Wu10,Ridolfo10,Manjavacas11}. They found that the quantum dot can compensate for the small absorption or scattering of the metallic particle and attributed the origin of these phenomena to the occurrence of Fano resonances. Here, we extend these pioneering works to unexplored physical regimes, where a quantum emitter enhances the absorption of a tiny gold nanoparticle by thousand times or cloaks a large metallic particle. We provide physical pictures for these fascinating phenomena by developing a general theoretical platform, where the interaction of an incident beam, a quantum emitter, and a metallic structure of arbitrary size, shape and material is treated using quantized electromagnetic fields. In addition, we point out shortcomings of some of the previous treatments and interpretations.

\textit{Theoretical approach} - Our formulation is based on the canonical quantization of the electromagnetic radiation in dispersive and absorptive media. This scheme has been developed by various authors~\cite{Huttner92,Matloob95} and was extended to inhomogeneous materials by Welsch and co-workers~\cite{Knoell01}. Here, the media are introduced as phenomenological noise currents in Maxwell's equations and the field operators are obtained indirectly from the noise operators via the classical Green's function. Furthermore, we include the effect of an external driving field and construct the system Hamiltonian,
\begin{align}\hat{H}&=\int d^3 r \int_{0}^{\infty}d\omega\hbar\omega\hat{f}^{\dagger} \left( r,\omega \right)f\left( r,\omega \right)+\frac{1}{2}\hbar\omega_{A}\hat{\sigma}_{z}\nonumber\\
&-\left(\mu \hat{E}^-\hat{\sigma}+\hat{\sigma}^{+}\hat{E}^{+}\mu \right),\end{align}
where $\hat{f}\left( r,\omega \right)$ and $\hat{f}^{\dagger}\left( r,\omega \right)$ are are the bosonic vector field operators for the elementary excitations of the system. The first term describes the electromagnetic field in the media, while the second term accounts for the energy of a two-level system (TLS) at transition frequency $\omega_{A}$ and population difference operator $\hat{\sigma}_{z}$. The third term considers the interaction between the emitter and the total electric field under the rotating wave approximation. Here $\mu$ is the transition dipole moment and $\hat{\sigma}$ is the coherence operator. The electric field operator $\hat{E}^{+}=E_{L}^{+}+\hat{E}_{v}^{+}+\hat{E}_{s}^{+}$ at the position ($r_{A}$) of the TLS consists of three contributions. The first component is the driving laser field ($E_{L}^{+}$) in the presence of the metallic structure. The second term is the vacuum field operator ($\hat{E}_{v}^{+}$) and the last part represents the quantized emission (source) field ($\hat{E}_{s}^{+}$). Since $E_{L}^{+}$ can be kept constant in the experiment, it is not a dynamic variable but a \textit{c}-number $E_{L}e^{-i\omega_{L}t}$, where $\omega_{L}$ is the driving frequency. The emission field can be expressed as~\cite{Knoell01},
\begin{equation}
\hat{E}_{s}^{+}=\int d^3r'\int_{0}^{\infty}\hspace{-.3cm}d\omega'\sqrt{\frac{\hbar\epsilon_{I}\left(r',\omega\right)}{\pi\epsilon_{0}}}\frac{i\omega'^2}{c^2}\textbf{G}\left( r,r',\omega'\right)\hat{f}\left(r',\omega' \right),
\end{equation}
where $c$ is the speed of light in vacuum, $\epsilon_{I}$ is the imaginary part of the dielectric constant and $\textbf{G}\left( r,r',\omega'\right)$ is the classical dyadic Green's function. Using the Markov approximation~\cite{Knoell01}, we obtain
\begin{align}
\hat{E}_{s}^{+}&=\frac{\omega_{L}^{2}}{\epsilon_{0}c^2}
\biggl[\textbf{P}\int_{0}^{\infty}d\omega \frac{\omega^2}{\pi \omega_{L}^{2}}\frac{\textbf{ImG}\left(r,r_{A},\omega \right)}{\omega-\omega_{L}}\nonumber\\
&+i\textbf{ImG}\left(r,r_{A},\omega_{L}\right)\biggr]\cdot\mu \hat{\sigma},
\end{align}
where \textbf{P} and \textbf{Im} stand for the principal value of the integral and the imaginary part of a function, respectively.
Next, we apply Eq.(3) to the Heisenberg equations of motion to arrive at
\begin{equation}
\dot{\hat{\sigma}}=-\frac{\gamma_{m}}{2}\hat{\sigma}-i\left(\omega_{A}+\delta\omega\right)\hat{\sigma}-i\frac{\Omega}{2}\hat{\sigma}_{z}e^{-i\omega_{L}t}-i\frac{\mu}{\hbar}\hat{\sigma}_{z}\hat{E}_{v}^{+}
\end{equation}
\begin{equation}
\dot{\hat{\sigma}}_{z}=i\left(\Omega\hat{\sigma}^+-\Omega^*\hat\sigma\right)-\gamma_{m}\left(1+\hat{\sigma}_{z}\right)
\end{equation}
with $\Omega=2\mu\cdot E_ {L}/\hbar$ as the complex Rabi frequency. The quantities $\gamma_{m}=\frac{2\omega_{L}^{2}}{\epsilon_{0}\hbar^2}\mu\cdot\textbf{ImG}\left(r_{A},r_{A},\omega_{L}\right)\cdot\mu$ represent the modified decay rate of the excited state and the frequency shift of the TLS transition in the presence of the metallic structure, respectively. For simplicity, we include $\delta\omega$ in $\omega_{A}$ as an implicit parameter and neglect it throughout the paper. Based on the above equations, it is straightforward to compute the expectation values of the operators and their various orders of correlation functions~\cite{Kimble76,Dorner02}. We, thus, obtain the Modified Optical Bloch Equations (MOBEs),
\begin{equation}\langle\dot{\hat{S}}\rangle=\left(-\frac{\gamma_{m}}{2}+i\delta_{L}\right)\langle\hat{S}\rangle-i\frac{\Omega}{2}\langle\hat{\sigma}_{z}\rangle,\end{equation}
\begin{equation} \langle\dot{\hat{\sigma}}_{z}\rangle=i\left(\Omega\langle\hat{S}^+\rangle-\Omega^*\langle\hat{S}\rangle\right)-\gamma_{m}\left(1+\langle\hat{\sigma}_{z}\rangle\right),\end{equation}
where $\hat{S}\left(t\right)=\hat{\sigma}\left(t\right)e^{i\omega_{L}}$, and $\delta_{L}=\omega_{L}-\omega_{A}$ is the frequency detuning. The MOBEs can be easily extended to deal with dephasing $\left(\gamma^*\right)$ and nonradiative decay$\left(\gamma_{nr}\right)$ rates if one replaces $\gamma_{m}$ by $\left(\gamma_{m}+\gamma^*+\gamma_{nr}\right)$ and $\left(\gamma_{m}+\gamma_{nr}\right)$ in Eqs.(6) and (7), respectively. By solving the steady-state equations, we obtain the expectation value of the induced dipole moment of the TLS as
\begin{equation} d=-\frac{\mu\Omega\left[2\delta_{L}-i\left(\gamma_{m}+\gamma^*+\gamma_{nr}\right)\right]}{4\delta_{L}^{2}+2\left|\Omega\right|^2\left(1+\frac{\gamma^*}{\gamma_{m}+\gamma_{nr}}\right)+\left(\gamma_{m}+\gamma^*+\gamma_{nr}\right)^2} \end{equation}
The expectation value of the total electric field operator becomes
\begin{equation} \langle\hat{E}\left(r\right)\rangle=E_{L}\left(r\right)+\frac{\omega_{L}^{2}}{\epsilon_{0}c^2}\textbf{G}\left(r,r_{A},\omega_{L}\right)\cdot d \end{equation}
Since all the geometrical and material information is included in the Green's function, the present method provides a general and rigorous treatment of the problem. Here, we emphasize that in deriving Eq. (4), we used the fact that the nonlinear term $\hat{\sigma}_{z}\left(t\right)\hat{\sigma}\left(t\right)$ collapses to $-\hat{\sigma}\left(t\right)$. This is a crucial difference between the present work and the semiclassical approaches applied in Refs.~\cite{Zhang06,Artuso08,Artuso11}, where this nonlinear term persists, giving rise to the peculiar nonlinear Fano effects. Moreover, it follows from Eq.(8) that for an ideal TLS (\textit{i.e.}$\gamma_{nr}=\gamma^*=0$), $d$ is independent of $\mu$ at weak excitation and on resonance. Therefore, contrary to some discussions in the literature~\cite{Artuso08,Manjavacas11,Artuso11},the size of the transition dipole moment is not a suitable measure for assessing the interaction strength.

In the following, we consider three regimes that lead to different phenomena. In regime I, the metallic particle is very small so that $\sigma_{0}\gg\sigma_{abs}\gg\sigma_{sca}$ while in regime II the particle is large and scattering-dominant such that $\sigma_{0}\ge\sigma_{sca}\gg\sigma_{abs}$. We shall see that the emitter and the metallic particle exchange their roles as optical antennas for each other in these cases. In regime III, we go beyond subwavelength particles and study metallic structures with $\sigma_{sca},\sigma_{abs}\ge\sigma_{0}$.

\begin{figure}[b!] \centering \includegraphics[width=8cm]{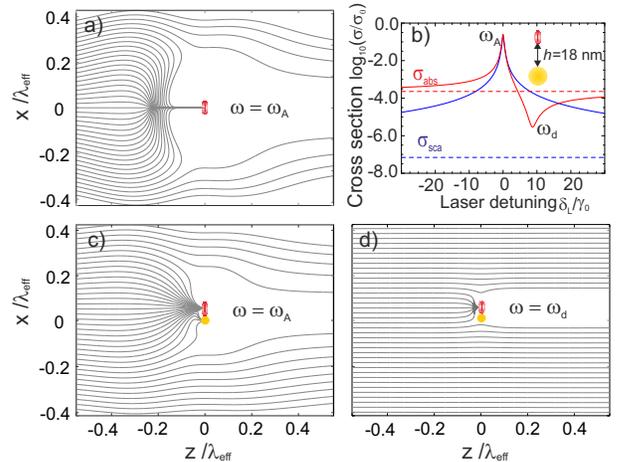} \caption{ a) Poynting flux lines flow from left to right for a TLS in a homogenous medium. b) Absorption and scattering cross sections of a composite system on a logarithmic scale. The dashed lines refer to the case of an isolated nanoparticle. The inset shows the arrangement of a gold nanosphere with diameter 5 nm at a distance of 18 nm from a quantum emitter oriented radially. c, d) Same as a) but for the composite system at $\omega_{A}$ (zero detuning ) and at $\omega_{d}$ }\label{smallMNP}
\end{figure}

\textit{Regime I} -  The Poynting flux lines in Fig. 1a illustrate that a naked TLS embedded in a medium of refractive index 1.5 scatters an incoming beam very effectively within an area $\sigma_{0}$~\cite{Paul83}. We now introduce a gold nanosphere with a diameter of 5 nm, a plasmon resonance at 545 nm, and scattering and absorption cross sections of $6.9\times10^{-8}\sigma_{0}$ and $2.3\times10^{-4}\sigma_{0}$, respectively. Such a gold particle has a negligible influence on the flux of an incident beam. To investigate the response of the coupled system to an incident light beam, we first calculate $\gamma_{m}$ and $E_{L}\left(r_{A}\right)$ using Mie theory~\cite{Chew99} and solve the steady-state MOBEs. Here we consider a plane wave with electric field along the TLS dipole moment ($\hat{x}$) and radial with respect to the nanosphere. We then evaluate the Poynting flux of the scattered field as well as the total field (c.f. Eq. (9)) through a surface enclosing the emitter-particle system. Fig. 1b shows the frequency dependence of the absorption and scattering cross sections of the composite (note the logarithmic scale) at a separation of $h=18nm$, where absorption reaches its maximum.

At zero detuning, both scattering and absorption cross sections experience several orders of magnitude increase. The Poynting flux lines in Fig. 1c are similar to the case of Fig. 1a and show that the system is dominated by the TLS and maintains a relatively large scattering cross section. One also sees, however, that some flux lines end at the particle where absorption can take place. Considering that $\sigma_{abs}$ for an isolated particle is as small as $2.3\times10^{-4}\sigma_{0}$  and that the net dissipation of a TLS is zero, it is remarkable that the absorption cross section of the combined system can become as large as $0.25\sigma_{0}$. The underlying physics of this intriguing phenomenon can be best understood by considering the TLS as an antenna that concentrates the propagating light in its vicinity: the strong scattering of the emitter leads to an enhanced intensity at the position of the gold particle and hence larger dissipation in the metal. We emphasize that this is different from a conventional quenching process, where the near-field excitation of higher-order multipoles extracts the energy out of the emitter in a nonradiative fashion. Here, the separation of the TLS and the gold particle is much larger than the particle radius so that the radiative and nonradiative rates of the emitter only amount to $1.2\gamma_{0}$ and$\gamma_{0}$, respectively, resulting in a quantum efficiency that is still high. We point out that previous reports ~\cite{Zhang06,Artuso08,Wu10,Ridolfo10,Manjavacas11} have not identified the parameter regime for such large enhancements of absorption.

At a frequency detuning of $\delta_{L}=8.75\gamma_{0}$, the absorption is minimized to $\sigma_{abs}=2\times10^{-6}\sigma_{0}$ (see Fig. 1b). The frequency-dependent change of absorption has been loosely attributed to Fano resonances in the literature~\cite{Ridolfo10,Manjavacas11}. Fano resonances entail constructive and destructive interference of two excitation paths ~\cite{Fano61}, which in this case would involve the TLS and the gold particle. However, considering that visible light penetrates metallic nanoparticles and that metals absorb at optical frequencies, one leg of such an interference process would involve dissipation. Thus, the observation that absorption can be made smaller than its intrinsic value of the naked particle appears to be puzzling. The Poynting flux lines in Fig. 1d shed light on this situation. We see that the beam avoids the metallic particle, hence minimizing dissipation. This can be again understood by considering the single quantum emitter as an antenna that attracts the propagating energy. It turns out that at detuning $\delta_{L}$ the field scattered from the TLS at the particle position has dropped enough to have the same magnitude but opposite sign as the incident field, so that a conventional destructive interference occurs for the driving field on the particle.

\begin{figure}[h!] \centering \includegraphics[width=8.5cm]{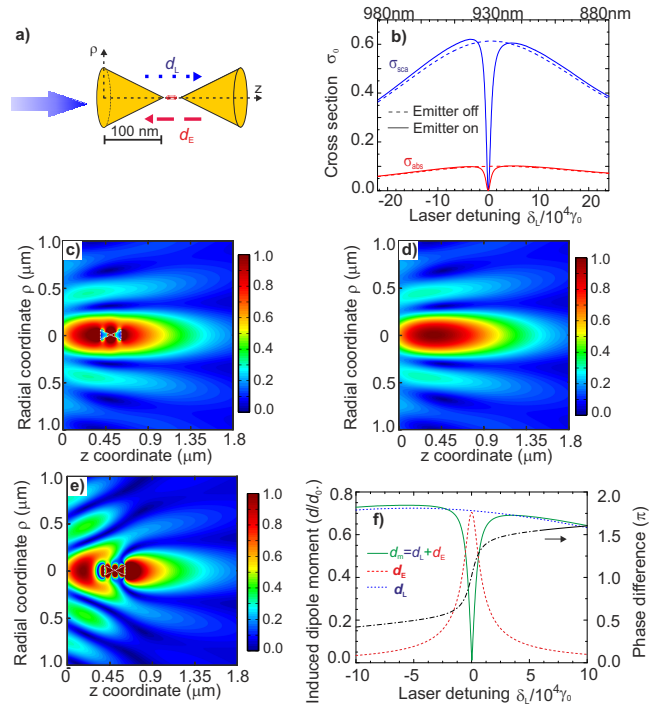} \caption{a) Schematics of the arrangement. The gold nanocones are 100 nm long, have a base diameter of 100 nm and are separated by 10 nm. The TLS lies in the mid gap with its dipole moment along the symmetry axis. b) Spectra of the absorption and scattering cross sections of the composite. c) The z-component electric field distribution of a FRPB propagating along z in the presence of the TLS and antenna. d) Same as c) but without the TLS or antenna. e) Same as c) but without TLS. f) Frequency dependence of the magnitudes of the induced dipole moments $d_{E}$, $d_{L}$, their sum ($d_{m}$) and their phase difference. Notation: dipole induced in TLS in homogenous medium ($d_{0}$), in the particle by the laser ($d_{L}$), in the particle by the TLS ($d_{E}$), in the metallic particle in total ($d_{m}$)} \end{figure}

\textit{Regime II} - We now replace the small gold cluster with a larger metallic structure that has a scattering-dominated extinction cross section comparable to $\sigma_{0}$. Figure 2a sketches the system consisting of a gold double-cone nanoantenna that sandwiches a TLS with its dipole moment oriented along the symmetry axis and is illuminated by a focused radially-polarized beam (FRPB)~\cite{Chen10}. This choice allows us to perform efficient calculations using body of revolution finite-difference time-domain method (BOR-FDTD)~\cite{Taflove05}. However, the physics discussed below remains valid if the system is driven by a plane wave with electric field polarized along the rotation axis. The TLS transition is set to coincide with the antenna plasmon resonance at 930 nm.

Figure 2b presents the striking finding that both absorption and scattering cross sections diminish on resonance, yielding an extinction dip as small as $0.001\sigma_{0}$ for the composite system. In other words, the combination of two strong scatterers, namely the plasmonic nanoantenna and the TLS, leads to transparency for the incoming light beam. Moreover, this happens over a broad spectral range of about 1THz because the double-cone nanoantenna enhances the spontaneous emission rate of the TLS by more than 11000 times ($\sigma_{0}=75$ MHz), while maintaining a quantum efficiency of 90\%~\cite{Mohammadi10} .

Figure 2c displays the z-component of the electric field when a FRPB illuminates the composite system. Comparison with the plots of Figs. 2d and 2e for the cases of a beam focused in free space and on a naked antenna (without TLS) helps visualize the large cloaking effect of a single quantum emitter on a metallic object of substantial cross section. It is clear that any scheme for controlling the emitter transition (e.g. Stark effect, photoswitching or optical pumping) allows active switching of the beam transmission. We have verified that this arrangement can also be extended to a macroscopic array of such structures, which would then switch the propagation of a plane wave. To this end, the composite system acts as an active metamaterial~\cite{Hess12}.

To gain physical insight into the situation at hand, we consider the dipole moments of the particle and the emitter as a function of frequency. The dipole moment induced in the emitter is directly calculated from Eq.(8) and scales roughly inversely with the square root of $\gamma_{m}$, becoming as weak as $7\times10^{-3}$ of its value in the absence of the antenna. The induced dipole moment of the metallic antenna can be computed according to $d_{m}=\epsilon_{0}\left(\epsilon_{Au}-\epsilon_{b}\right)\int d^3r E$, where $E$ is the total electric field in the metal obtained from BOR-FDTD (c.f. Eq.(9)). The green trace in Fig. 2f shows that $d_{m}$ is reduced on the TLS resonance. This reduction results from the destructive interference of the dipoles induced in the antenna by the direct excitation field ($d_{L}$) and by the TLS ($d_{E}$). The phase of the antenna near field at the TLS position is retarded by $90^{\circ}$ with respect to the incident light. Because the excitation of the TLS is dominated by this near field and the field scattered by the emitter is again retarded by $90^{\circ}$ on its resonance, $d_{E}$ is out of phase with $d_{L}$. Figure 2f summarizes the magnitude and phase difference of the relevant dipole moments as a function of the laser detuning.

The reduced polarization of the antenna also explains the lower absorption (see Fig. 2b). We note that optimization of parameters (antenna geometry, TLS position, etc.) for a large enhancement of the spontaneous emission plays a key role because it signifies the strength of the coupling between the emitter and the antenna. The underlying physics of this regime is analogous to the recent findings, where the couplings of a dipole and a resonator~\cite{Waks06} or of two resonators ~\cite{Smith04} lead to transparency. In all these cases, it is the conventional interference of electromagnetic fields that gives rise to transparency.

\begin{figure}[t h!] \centering \includegraphics[width=7.5cm]{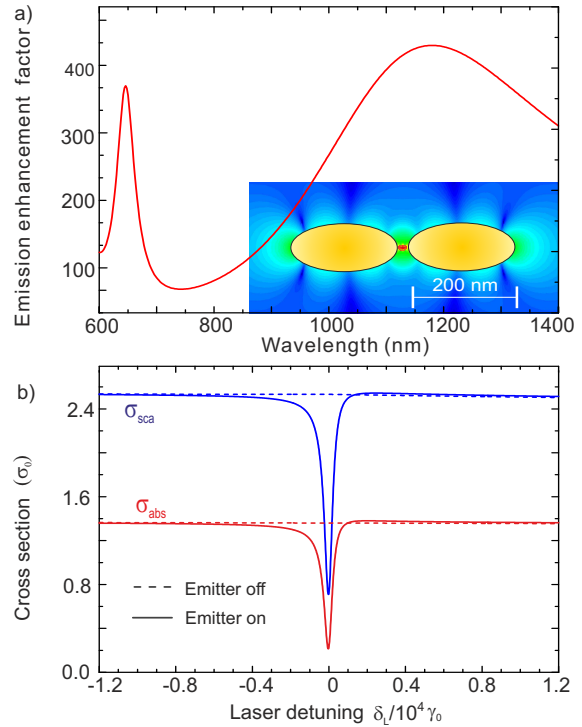} \caption{a) Enhancement of the spontaneous emission rate as a function of the emitter transition wavelength. The TLS was placed in the 20 nm gap between two gold prolate spheroids with long and short axes of 200 nm and 100 nm, respectively. Inset plots the field distribution of a dipole radiating at octupolar resonance 642 nm. b) Spectra of the absorption and scattering cross sections of a TLS coupled to the antenna}\end{figure}

\textit{Regime III} - In the previous sections, we studied subwavelength metallic structures at dipolar plasmon resonances. Next, we examine the coherent interaction of a quantum emitter with larger metallic structures that support multipolar resonances and satisfy $\sigma_{sca}\ge\sigma_{abs}\ge\sigma_{0}$. As an example, we consider a pair of gold prolate spheroids with dimensions of about $\lambda_{eff}=\lambda/\sqrt{\epsilon_{b}}$ under illumination by a FRPB (see the inset of Fig. 3a). The spectrum in Fig. 3a displays the wavelength dependence of the TLS spontaneous emission rate, revealing enhancements at the dipolar (multipole orde $l=1, \lambda=1180 nm$) and octupolar ($l=3, \lambda=642nm$) resonances. We note that although the transition of the TLS is dipolar, it can excite higher order modes of the antenna, which contrary to the case of small particles, can radiate efficiently~\cite{Mertens07}. However, symmetry arguments in our geometry prevent the emitter from coupling to even multipole orders such as $l=2$ (quadrupole).

The behavior of the system in the dipolar region is quite similar to the previous regime so we focus our attention on a TLS with a transition at 642 nm. As displayed by the dashed lines in Fig. 3b, scattering and absorption of the bare metallic structure amount to an extinction cross section of $3.9\sigma_0$ near the octupolar resonance. The solid lines in Fig. 3b show that also in this case the presence of a single TLS can strongly counteract both the scattering and absorption of the metal. In fact, the reason for an incomplete transparency is that the incident light does not fully couple to the dipolar or octupolar modes.

\textit{Conclusion} - In this Letter, we investigated the scattering and absorption of composite materials made of quantum emitters and plasmonic structures with various dimensions. We found that an emitter acts as an optical antenna for very small particles to either reduce or enhance their absorption by thousand times. Large absorption has a number of immediate applications, for example, in the realization of efficient nanoscopic heat sources ~\cite{Govorov06} or high-efficiency generation of hot electrons ~\cite{Knight11}. We also demonstrated that an emitter could cloak a strong scatterer in a frequency range determined by the enhanced spontaneous emission rate. An interesting application of this effect could be in the realization of single-photon power limiters, whereby saturation of the TLS destroys the transparency and causes attenuation of a single-photon stream.

The possibility of switching the scattering and absorption of passive metallic nanostructures with a single emitter has far-reaching implications for the control of signals in quantum networks and metamaterials. The theoretical platform presented here can be readily applied to the coherent excitation of emitters in the near field of surfaces and nanowires ~\cite{Gonzalez-Tudela11,Chang06,Dzsotjan11}. Furthermore, it could be extended to consider time-dependent events and pulsed excitation as well as gain and stimulated emission ~\cite{Hess12}.

We acknowledge financial support from the Max Planck Society, the ERC advanced grant SINGLEION, the Alexander von Humboldt Foundation and the EU-STREP project ``QIBEC''.  X.W. C. would like to thank Gert Zumofen for inspiring discussions.

%merlin.mbs 2010-03-15 4.21a (PWD, AO, DPC)

%\bibliography{vahid}
\end{document}